\documentclass[10pt,showpacs,amsmath,amssymb,floatfix,superscriptaddress,longbibliography,twocolumn,eqsecnum]{revtex4-2}
\usepackage{booktabs}
\usepackage{amsmath}
\usepackage{physics}
\usepackage{graphicx}
\usepackage[usenames,dvipsnames,svgnames,table]{xcolor}
\usepackage{tcolorbox}
\counterwithout{equation}{section}
\counterwithout{figure}{section}
\usepackage[unicode=true,
            pdfusetitle,
            bookmarks=true,
            bookmarksnumbered=false,
            bookmarksopen=false,
            breaklinks=true,
            pdfborder={0 0 0},
            backref=false,
            colorlinks=true,
            hypertexnames=false]{hyperref}
\hypersetup{linkcolor=NavyBlue,urlcolor=NavyBlue,citecolor=NavyBlue}

\begin{document}

\title{Multi-modes Bessel-Gaussian-Orbital Angular Momentum Beams Quantum Holography}
\author{Jinjin Li}
\affiliation{%
College of Physics, Hangzhou Dianzi University, Zhejiang 310018, China.
}%

\author{Chaoying Zhao}%
\email{Corresponding author: zchy49@163.com}
\affiliation{%
College of Physics, Hangzhou Dianzi University, Zhejiang 310018, China.
}%
\affiliation{%
State Key Laboratory of Quantum Optics and Quantum Optics Devices, Institute of Opto-Electronics, Shanxi University, Taiyuan 030006, China
}%
\affiliation{%
Zhejiang Key Laboratory of Quantum State Control and Optical Field Manipulation, Hangzhou Dianzi University, Hangzhou 310018, China
}%

\date{\today}

\begin{abstract}
We propose an orbital angular momentum (OAM) quantum holography scheme based on multi-mode Bessel-Gaussian (MBG) beams. Entangled photon pairs are generated through spontaneous parametric down-conversion (SPDC) process, and the axis prism parameters and topological charges of the idler photons are used for encoding to construct Bessel-Gaussian quantum selective holograms; then, the corresponding mode parameters carried by the signal photons are used for correlated decoding and information reconstruction. Theoretical analysis and numerical simulation results show that this scheme can effectively realize OAM quantum holography based on Bessel-Gaussian modes encoding. Compared with traditional single OAM encoding methods, our scheme introduce an additional mode degrees of freedom, which can enhance  multiplexing dimension and encoding capacity; at the same time, relying on the non-classical correlation characteristics of entangled photons, quantum holography has a potential advantages in noise-resistance performance.
\end{abstract}

\maketitle

\section{Intrdouction}

In 1948, Dennis Gabor \cite{1} introduced a holography technique to reconstruct three-dimensional light field information of an object, by capturing the interference pattern between reflected light and reference light enabling the recreation of virtual objects. Since its invention, it has been extensively utilized in three-dimensional holographic display \cite{2}, data storage \cite{3}, optical encryption \cite{4}, holographic interferometry \cite{5}, and microscopy \cite{6}. We employ the physical properties of light, such as wavelength \cite{7}, polarization \cite{8}, spatial \cite{9}, and temporal dimensions \cite{10}, to multiplex several parallel information channels within a single hologram. In 1992, Allen et. al. found \cite{11} that Laguerre-Gaussian (LG) beams with orbital angular momentum (OAM), characterized by a helical phase factor $e^{il\phi}$, where $l$ denotes the topological charge, which can potentially assume any integer or fractional value. It has an unlimited number of eigen-states, each associated with a distinct topological charge, and the eigen-states are orthogonal to one another. This enables a single photon to theoretically transmit an endless quantity of information bits. Consequently, OAM can serve as a spatial degree of freedom for encoding a photon.

The swift advancement of computers has facilitated the rise of computational holography. Employing computational holography techniques to supplant conventional holography eliminates the necessity for establishing intricate interference optical pathways. Nonetheless, typical optical modulation systems are limited by challenges including low resolution, substantial size, and a singular modulation dimension \cite{12,13,14}. In 2019, Ren et. al.\cite{15} initially implemented the principle of meta-surface orbital angular momentum holography, utilizing gallium nitride nano-pillars to construct meta-surface OAM holographic devices. By sampling in the spatial frequency domain, they circumvented the overlap of OAM helical wave-fronts, resolving the deficiency of OAM selectivity in conventional hologram design and accomplishing the multiplexing and demultiplexing of OAM holography. The integration of the spatial dimension of OAM with meta-surface holography enables OAM-selective holography via meticulous regulation of the meta-surface's amplitude and phase. The target image can only be deciphered in the presence of an incident OAM beam with a specified topological charge, thereby considerably augmenting information security.

Simultaneous acquisition of large-capacity and high-resolution holograms is unattainable, as an increase in the information content of an OAM hologram results in diminished resolution, and crosstalk frequently arises between distinct OAM channels of the hologram. To preserve the OAM properties at the sampling spots, we can conduct sparse sampling on the original image. Nevertheless, as the quantity of modes escalates, the sampling constant correspondingly rises, ultimately rendering the image indistinguishable \cite{16,17,18,19,20}. In 2024, Ji et. al.\cite{21} introduced a Bessel-Gaussian (BG) beam OAM holography system that significantly mitigates the constraints of sample constants for various OAM modes. Utilizing the self-healing ability of BG beams during propagation enables the generation of a perfect optical vortex (POV) \cite{22} via Fourier processing, with distinct modes of POV exhibiting a stable ring radius, hence mitigating the constraints of the sampling constant. Multi-mode vortex light denotes a superposition of orbital angular momentum beams with varying topological charges \cite{23}, and conducting holography on multi-mode vortex light can significantly enhance the quality of holograms. Multi-mode vortex holography is accomplished by varying the combination of multi-mode vortex light to modify the superimposed phase, which is the aggregate of the phases of separate single-mode OAM beams, thereby creating encoded holograms with diverse decoding techniques. In recent years, high-capacity OAM holography technology has been integrated with quantum entangled light sources, achieving holographic technology with high-dimensional entanglement capabilities. In 2023, a research team \cite{24} introduced a high-dimensional quantum holography scheme by utilizing high-dimensional OAM-entangled photons. This approach employs OAM-selective holographic encoding and projection measurement to realize high-dimensional quantum holography and multi-channel holographic multiplexing. In comparison to classical holography, quantum holography has numerous advantages, such as enhanced resistance to classical noise and random phase interference, as well as improved resolution.

Based on the above discussion, we propose a high-dimensional quantum holography theory based on the OAM modes of Bessel-Gauss beams, and we verify the correctness of the MBG-OAM quantum holography scheme through simulation experiments, confirming that our scheme has stronger noise resistance compared to classical schemes. First, we generate OAM-entangled photon pairs via spontaneous parametric down-conversion and encode the idler photon path using axicon parameters and topological charge to construct the Bessel-Gaussian quantum selective hologram; then we perform correlated decoding and information reconstruction based on the corresponding mode parameters carried by the signal photons. Theoretical analysis and numerical simulation results show that images can be reconstructed in single-mode, dual-mode, and four-channel multiplexing scenarios. Finally, we objectively evaluate the differences between images and quantify image quality using mean square error (MSE), peak signal-to-noise ratio (PSNR), and structural similarity index (SSIM), and we analyze the PSNR for dual-mode and four-channel multiplexing both with and without noise, confirming the noise resistance capability of our proposed MBG-OAM quantum holography scheme.

\section{Principles and methods}
\subsection{MBG-OAM Quantum Selective Hologram}

We design MBG-OAM selection holography based on the GS algorithm, as shown in Fig.1.
\begin{figure}[htp]
	\centering
\includegraphics[width=0.9\linewidth]{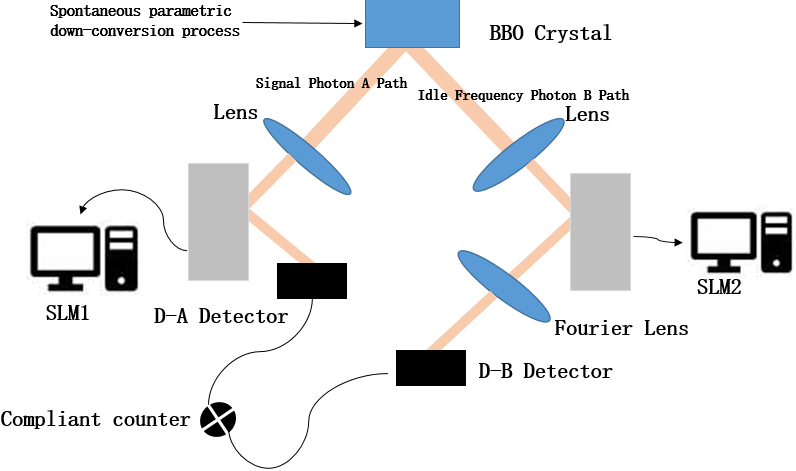}
	\caption{MBG-OAM Quantum Selective Holographic Diagram.}
\end{figure}

Based on spontaneous parametric down-conversion (SPDC) technology, pump light passing through a BBO crystal to generate OAM entangled photon pairs. Their OAM spiral bandwidth is shown in Fig.2. \cite{25}\\
\begin{figure}[htp]
	\centering	\includegraphics[width=0.8\linewidth]{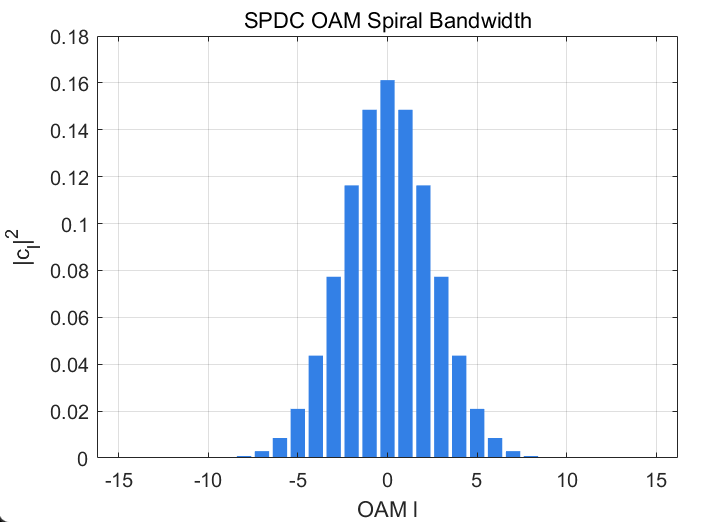}
	\caption{OAM Spiral Bandwidth in the SPDC Process.}
\end{figure}

Specifically, the signal photon $s$ undergoes OAM mode decoding through SLM1, while the idler photon $i$ has the phase of the MBG-OAM quantum holography loaded on SLM2. After passing through a Fourier lens, scanning detection is performed in the Fourier plane, and coincidence counting measurement is carried out through D-A and D-B, thus enabling the reconstruction of the quantum holographic image. The Gaussian point image of the target can only be reconstructed when the phases of the MBG on SLM1 and SLM2 are opposite.

The OAM entangled state can be expressed by
\begin{equation}
	\left |\hat{\Phi}\right \rangle = {\textstyle\sum_{l=-\infty }^{l=+\infty }} \hat{a}_{l}\left|l \right \rangle_{s}\left|-l \right \rangle_{i},  
\end{equation}
Here, the subscript $s$ denotes the signal photon, $i$ denotes the idler photon, $l$ denotes the topological charge, and $a$ denotes the axicon parameter. In this MBG-OAM quantum holographic scheme, the axis prism parameter $a$ and the topological charge $l$ are used to encode the state, where $a$ determines the state $\left|l\right\rangle$ spatial probability distribution radius, the value of $a$ of each spatial mode should be close, if the difference of $a$ is too large, which will lead to $ |l_{i}\left\rangle|l_{j}\right\rangle$. The overlap between the spatial modes will decrease, which will affect the pattern matching and the projection measurement efficiency. The selective projection of the MBG-OAM quantum space mode is realized through spatial aperture adjustment to ensure that the multi-mode superposition state maintains a good pattern matching and a high detection efficiency during propagation and measurement.
We use the two-photon correlation matrix to characterizing the joint distribution relationship of the signal photon and the idler photon in the mode space, as shown in Fig.3. 
\begin{figure}[htp]
	\centering
\includegraphics[width=1\linewidth]{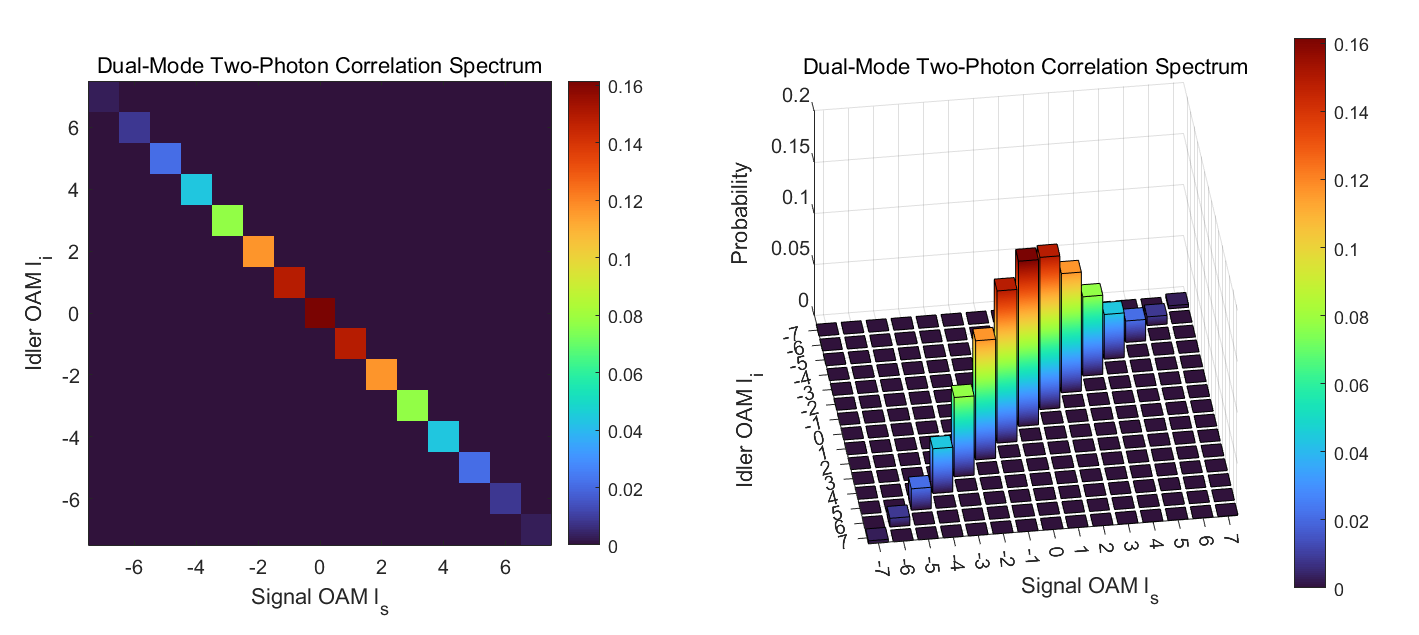}
	\caption{Two-photon correlation matrix.}
\end{figure}

\subsection{Two-photon MBG state and superposition state}
The spatially normalized lateral mode function of MBG single photons is defined as
\begin{equation}
	u_{l_{j},\kappa}(r,\varphi,z)
	= \mathcal{N}_{l_{j},\kappa} J_{l_{j}}(\kappa r)e^{i l_{j} \varphi} 
	e^{-r^2/w_0^2}e^{i k_z z},
\end{equation}
where $w_0$ represents the Gaussian beam waist radius, $\kappa \equiv k_r$ represents the transverse wave-vector parameter, $\mathcal{N}_{l_{j},\kappa}$ represents the normalized constant, which satisfy in cross-section
\begin{equation}
\int_{0}^{2\pi}\int_{0}^{\infty} 
\left|u_{l_{j},\kappa}(r,\varphi,z) \right|^{2} rdr d\varphi = 1.
\end{equation}
and $J_{l_{j}}(\kappa r)$ is represented by the first class Bessel function of the $l$ order.
\begin{equation}
	J_{l_{j}}(\kappa r)=\sum_{j=0}^{\infty} \frac{(-1)^{j}}{j!\Gamma(j+l+1)}\left(\frac{\kappa r}{2}\right)^{2 j+l},
\end{equation}
with
\begin{equation}
	\Gamma(j+l+1)=(j+l)!,
\end{equation}

Based on the MBG single photon, we define the spatially normalized transverse mode function of the MBG two photons as
\begin{equation}
	\Phi(\mathbf{r}_s,\mathbf{r}_i)
	= \sum_{l_s,a_s} \sum_{l_i,a_i}
	C_{l_s,a_s; l_i,a_i}\,
	u^{(s)}_{l_s,a_s}(\mathbf{r}_s)\,
	u^{(i)}_{l_i,a_i}(\mathbf{r}_i),
\end{equation}

where $u^{(s)}_{l_s,a_s}(\mathbf{r}_s)$ represents the spatially normalized lateral mode function of signal photons, $u^{(i)}_{l_i,a_i}(\mathbf{r}_i)$ represents the spatially normalized lateral mode function of idle frequency photons, $C_{l_s,a_s;l_i,a_i}$ is the two-photon joint expansion coefficient.

Considering that the frequency is $\omega$, under the condition of approximation of the paraxial axis, the positive frequency part of the electric field operator of the signal photon channel is expanded by MBG mode as follows:
\begin{equation}
	\hat{E}_s^{(+)}(\mathbf{r}_s, t_s)
	= j \sum_{l_s, a_s} \sqrt{\frac{\hbar \omega_s}{2 \varepsilon_0}}
	\, u^{(s)}_{l_s, a_s}(\mathbf{r}_s)
	\, e^{-i \omega_s t_s}
	\, \hat{a}_{s; l_s, a_s},
\end{equation}

The positive frequency part of the idle-mode photon channel electric field operator is expanded in the MBG mode as:
\begin{equation}
	\hat{E}_i^{(+)}(\mathbf{r}_i, t_i)
	= j \sum_{l_i, a_i} \sqrt{\frac{\hbar \omega_i}{2 \varepsilon_0}}
	\, u^{(i)}_{l_i, a_i}(\mathbf{r}_i)
	\, e^{-i \omega_i t_i}
	\, \hat{a}_{i; l_i, a_i},
\end{equation}

where $\hat{a}_{s; l_s, a_s}$ and $\hat{a}_{i; l_i, a_i}$ indicates the annihilation operator.
$\hat{a}_{s; l_s, a_s}^{\dagger}$ and $\hat{a}_{i; l_i, a_i}^{\dagger}$ indicates the creation  operator, and $\omega$ represents the angular frequency of the light field.

The Hermitian conjugate of the positive frequency part is expressed as the negative frequency part:
\begin{align}
	\hat{E}_s^{(-)}(\mathbf{r}_s, t_s)
	&= \left[\hat{E}_s^{(+)}(\mathbf{r}_s, t_s) \right]^{\dagger}, \\[4pt]
	\hat{E}_i^{(-)}(\mathbf{r}_i, t_i)
	&= \left[\hat{E}_i^{(+)}(\mathbf{r}_i, t_i) \right]^{\dagger}. 
\end{align}

Therefore, we can represent the two-photon MBG-OAM mode state as:
\begin{equation}
	|\hat{1}_{l_s,a_s}\rangle_s \, |\hat{1}_{l_i,a_i}\rangle_i
	= \hat{a}_{s; l_s, a_s}^{\dagger} \, \hat{a}_{i; l_i, a_i}^{\dagger} |0\rangle,
\end{equation}
For the superposition state of the MBG mode, it is expressed as:
\begin{equation}
	|\hat{\Phi}\rangle
	= \sum_{l_s, a_s} \sum_{l_i, a_i}
	C_{l_s, a_s; l_i, a_i}
	\, |\hat{1}_{l_s, a_s}\rangle_s \, |\hat{1}_{l_i, a_i}\rangle_i,
\end{equation}
The probability amplitude of two photons in space can be expressed as:
\begin{equation}
	\Phi(\mathbf{r}_s, \mathbf{r}_i)
	= \langle 0| 
	\hat{E}_i^{(+)}(\mathbf{r}_i)
	\hat{E}_s^{(+)}(\mathbf{r}_s)
	|\hat{\Phi}\rangle,
\end{equation}

Its probability distribution satisfies:
\begin{gather}
\begin{aligned}
	P(\mathbf{r}_s, \mathbf{r}_i)
	\propto
	\langle \hat{\Phi} |
	\hat{E}_s^{(-)}(\mathbf{r}_s)
	\hat{E}_i^{(-)}(\mathbf{r}_i)
	\hat{E}_i^{(+)}(\mathbf{r}_i)
	\hat{E}_s^{(+)}(\mathbf{r}_s)
	|\hat{\Phi} \rangle\\
	= |\Phi(\mathbf{r}_s, \mathbf{r}_i)|^2,
    \end{aligned}
\end{gather}

If the two-photon state takes the form of an equal-weight superposition of two sets of MBG modes:
\begin{equation}
	|\hat{\Phi}\rangle = 
	\frac{1}{\sqrt{2}}
	\left(
	|\hat{1}_{l_{s1}, a_{s1}}\rangle_s |\hat{1}_{l_{i1}, a_{i1}}\rangle_i
	+ e^{j\delta}
	|\hat{1}_{l_{s2}, a_{s2}}\rangle_s |\hat{1}_{l_{i2}, a_{i2}}\rangle_i
	\right),
\end{equation}

where $\delta$ is the relative phase of the two-photon mode superposition.

Then interference terms related to the angles appear in the joint probability distribution:
\begin{equation}
	(l_{s1}-l_{s2})\varphi_s+(l_{i1}-l_{i2})\varphi_i.
\end{equation}

If the OAM conservation condition $l_s+l_i=l_p$ is further satisfied,
the interference structure can be further simplified.
For the OAM pump of 0, its angular modulation depends on $\varphi_s-\varphi_i$.

\section{Simulation and Analysis}
\subsection{Single-Modal State Reconstruction}
First, we examine the MBG state of a set of parameters. The process of encoding and decoding by using a set of parameters-axial prism parameter $a$ and topological charge number $l$ is referred to as single-mode OAM quantum holography. The parameters $(-a,-l)$ involved in encoding for specific idle-frequency photons. When we reconstruct the image, the parameters of the signal photons must satisfy $(a,l)$. The reconstruction results are shown in Fig.4. 
\begin{figure}[htp]
	\centering
\includegraphics[width=1\linewidth]{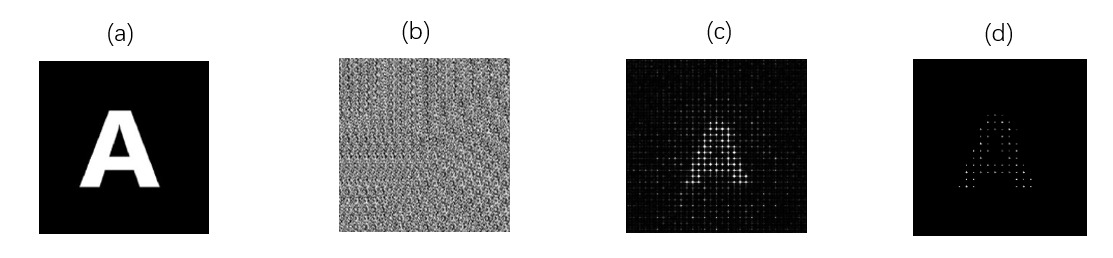}
	\caption{Single-mode reconstruction result images: (a)target image, (b)selected hologram, (c)image during unfiltered reconstruction, (d)image after filtered reconstruction.}
\end{figure}

\subsection{Dual-Mode Superposition State Reconstruction}
Correspondingly, the process of encoding and decoding by using two sets of parameters-axial prism parameter $a$ and topological charge $l$ is referred to as dual-mode OAM quantum holography. In order to verify the feasibility of the MBG-OAM quantum holography scheme, we simulate the phase of MBG-OAM quantum holography on idle-frequency photons and test the MBG states with two parameters. The entire simulation process was implemented by using computational holography. The encoding parameters are represented as $(-a_1,-l_1;-a_2,-l_2;\dots)$ for specific signal photons. When we reconstruct the image, the parameters of the signal photons must satisfy $(a_1,l_1;a_2,l_2;\dots )$.

First, we sample the original image to obtain the MBG-OAM quantum-preserved hologram. Then, after encoding the idle-frequency photons with $(-0.06,-3;-0.08,-2)$, we obtain the MBG-OAM quantum-selected hologram. We use the parameters $(0.06,3;0.08,2)$, $(0.06,3)$, $(0.08,2)$, $(0.08,3)$, $(0.12,4;0.12,1)$ and $(0.06,3;0.08,2;0.2,5)$ for separate encoding, and then decode them through the signal photons to obtain the results shown in Fig.5:
\begin{figure}[htp]
	\centering
	\includegraphics[width=1\linewidth]{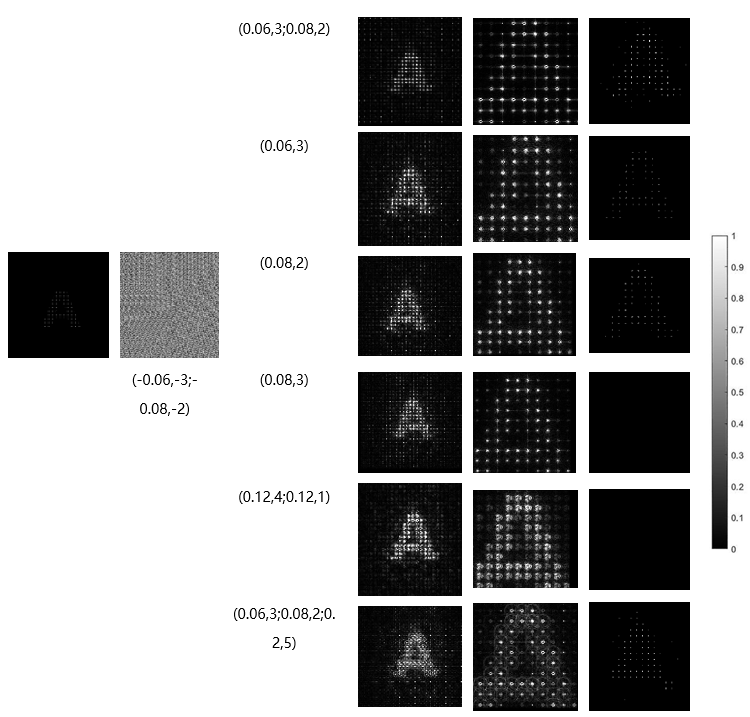}
	\caption{Dual-mode reconstruction images:  target image, (b) selected hologram and its parameters, (c) unfiltered reconstruction image, (d) detail image, (e) filtered reconstruction image.}
\end{figure}

The above results indicate that the image is only reconstructed when the topological charge and axicon parameters cancel each other out. When the signal photons are encoded as $(0.06,3;0.08,2)$, $(0.06,3)$, $(0.08,2)$, or $(0.06,3;0.08,2;0.2,5)$, the image is reconstructed. However, for other parameters such as $(0.08,3)$ and $(0.12,4;0.12,1)$, there are no topological charges and axicon parameters to cancel each other out, so the image is not reconstructed.

\subsection{Multiplexed Hologram Reconstruction}
Using the combination of axicon prism parameters and topological charge for multiplexing holograms further improves image imaging quality. The MBG-OAM multiplexed holograms formed by different coding combinations gain more coding degrees of freedom through the inclusion of axicon prism parameters, greatly enhancing the multiplexing channels and the security of decoding. To verify the feasibility of our quantum multiplexed holography scheme, we simulated multi-channel MBG-OAM multiplexed holograms. First, we sampled the original image “A” to obtain an OAM quantum-preserved hologram. Using idle-frequency photon coding as $(-0.06,-3;-0.08,-2)$, we obtain quantum selective hologram 1. Next, we sample the original image “B” to obtain an OAM quantum-preserved hologram. Using idle-frequency photon coding as $(-0.1,-8;-0.1,-10)$, we obtain quantum selective hologram 2. Then, we sample the original image “C” to obtain an OAM quantum-preserved hologram. Using idle-frequency photon coding as $(-0.02,-15;-0.02,-18;-0.02,-20)$, we obtain quantum selective hologram 3. Finally, we sample the original image “D” to obtain an OAM quantum-preserved hologram. Using idle-frequency photon coding as $(-0.1,-5;-0.02,-20)$, we obtain quantum selective hologram 4. By superimposing the four selective holograms, a multiplexed hologram was created, with the positions of the four target images in the hologram non-overlapping. The signal photons are respectively coded as $(0.02,18)$, $(0.02,20)$, $(0.06,3; 0.06,2)$, and $(0.1,10)$ to reconstruct the images. The reconstruction results are shown in Fig.6:
\begin{figure}[htp]
	\centering
	\includegraphics[width=1\linewidth]{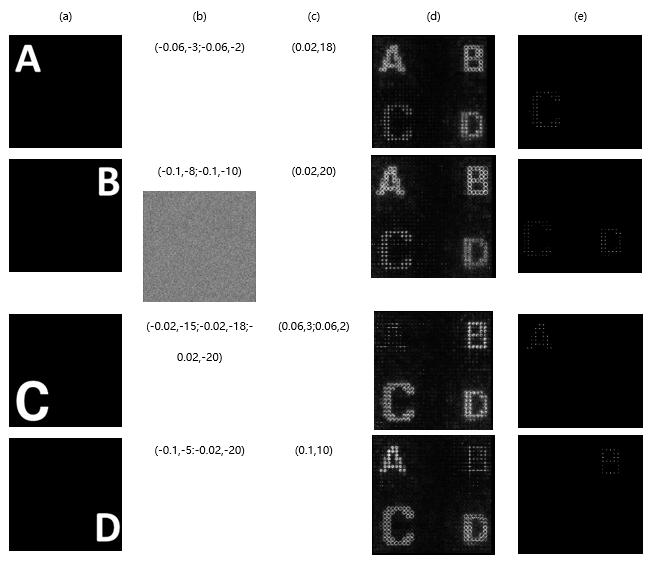}
	\caption{Reusable hologram reconstruction images: (a) Image information before sampling, (b) Reusable hologram and parameter combination, (c) Combined parameters, (d) Unfiltered reconstruction image, (e) Filtered reconstruction image.
	}
\end{figure}
(b) in Fig.7 is a detailed view of the multiplexed hologram in Fig.6(b). (c) in Fig.7 is a local detail view of Fig.6(e). The four non-overlapping regions (a) in Fig.7 are encoded as\textcircled{1}, \textcircled{2}, \textcircled{3}, and\textcircled{4}. As shown in the first image of the first column in Fig.6(e), the information “C” at position \textcircled{3} is reconstructed after filtering, as shown in the first image of Fig.7(c), while the information “A”, “B”, and “D” at positions\textcircled{1}, \textcircled{2}, and \textcircled{4} cannot be reconstructed after filtering. The second image in Fig.7(c) is a detail view of positions \textcircled{3} and \textcircled{4} in the second image of Fig.6(e); the third image in Fig.7(c) is a detail view of position \textcircled{1} in the third image of Fig.6(e); and the fourth image in Fig.7(c) is a detail view of position \textcircled{2} in the fourth image of Fig.6(e). Except for the above positions, the rest of the positions contain no information reconstruction after filtering.
\begin{figure}[htp]
	\centering
	\includegraphics[width=1\linewidth]{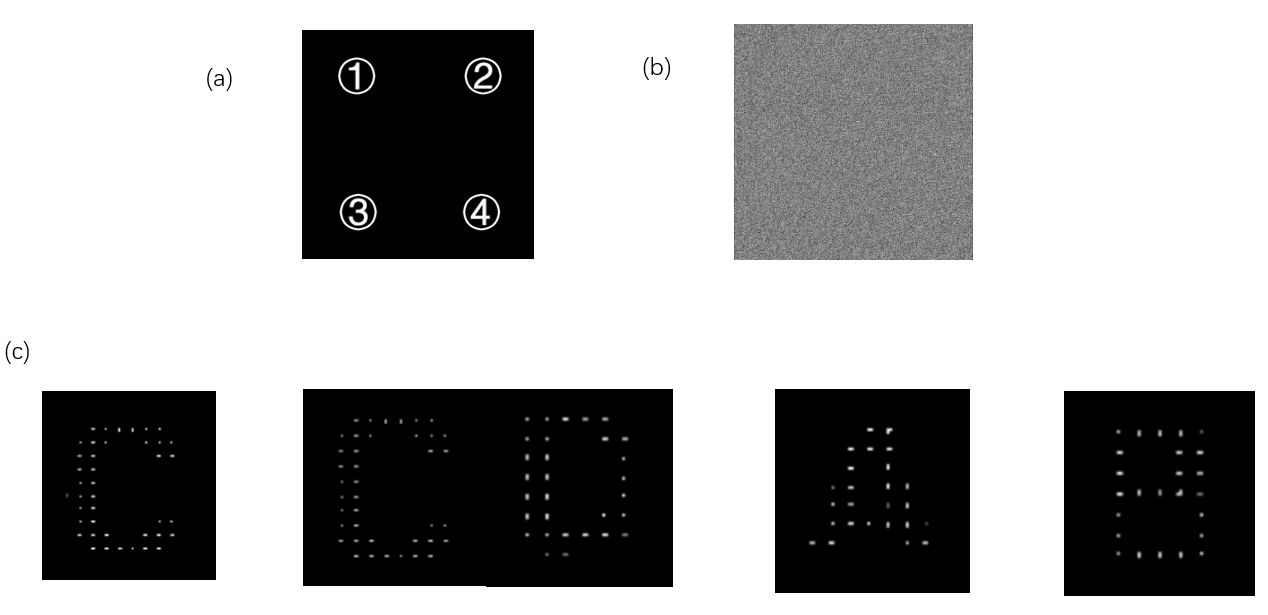}
	\caption{Reusable hologram reconstruction result images: (a) Image information before sampling, (b) Reusable hologram and parameter combination, (c) Combined parameters, (d) Unfiltered reconstructed image, (e) Filtered image.
	}
\end{figure}
According to the reconstruction results shown in Fig.6, when the combined parameters are set to $(0.02,18)$, $(0.02,20)$, $(0.06,3; 0.06,2)$, and $(0.1,10) $, the information of C, CD, A, and B is reconstructed. The above simulation results indicate that the quantum holographic multiplexing scheme we proposed is feasible.

\section{ANAYLYSIS OF IMAGE QUALITY IN LOSS AND RECONSTRUCTION}
\begin{figure}[htp]
	\centering
	\includegraphics[width=1\linewidth]{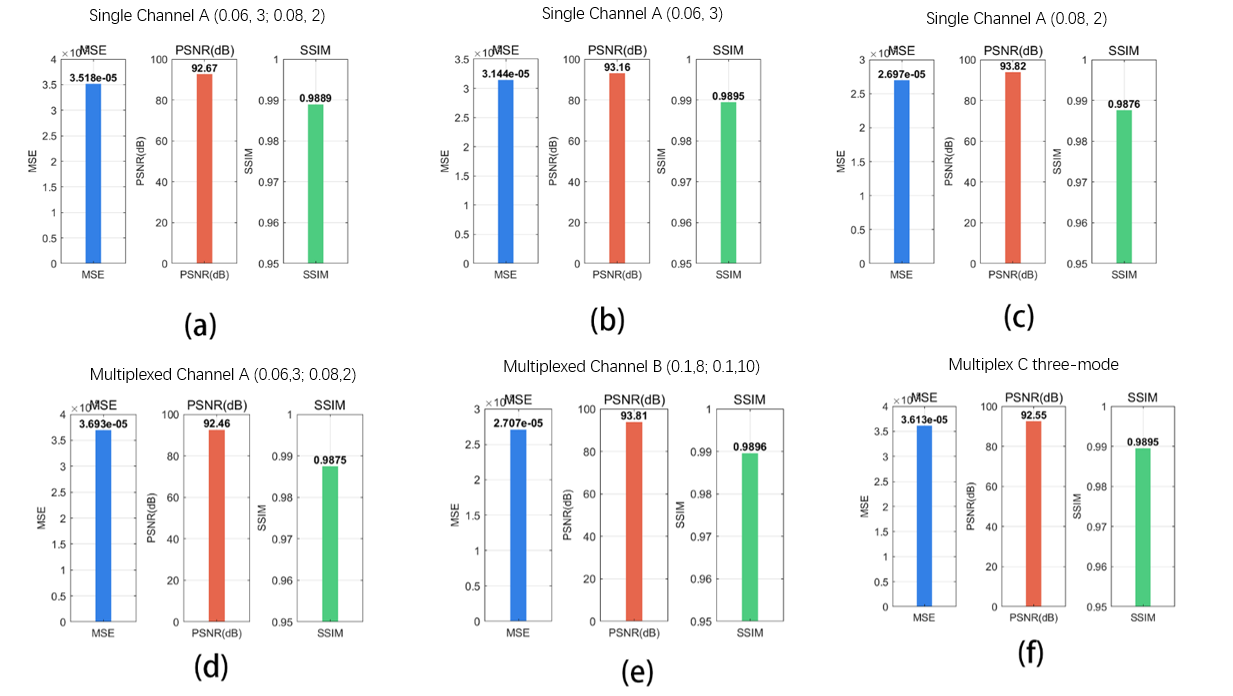}
	\caption{Mean Squared Error (MSE), Peak Signal-to-Noise Ratio (PSNR), Structural Similarity Index (SSIM) under each mode}
\end{figure}
In order to evaluating the quality of the outputs of this scheme, we use three indices-Mean Squared Error (MSE), Peak Signal-to-Noise Ratio (PSNR), and Structural Similarity Index (SSIM)-to objectively assess the differences between images and quantify image quality, as shown in Fig.8.
\begin{figure}[htp]
	\centering
\includegraphics[width=1\linewidth]{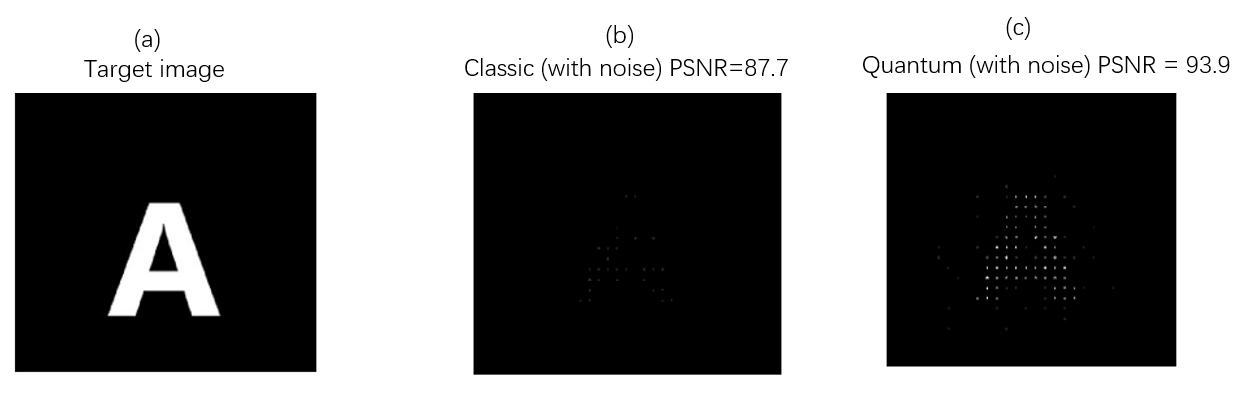}
	\caption{Classical and quantum peak signal-to-noise ratio (PSNR) in noisy conditions}
\end{figure}

According to the classical and quantum PSNR in noisy conditions as shown in Fig.9, we can see that the classical PSNR is $87.7 dB$, while the quantum PSNR is $93.9 dB$, which indicates that the quantum method has a stronger noise resistance.
\begin{figure}[htp]
	\centering
\includegraphics[width=1\linewidth]{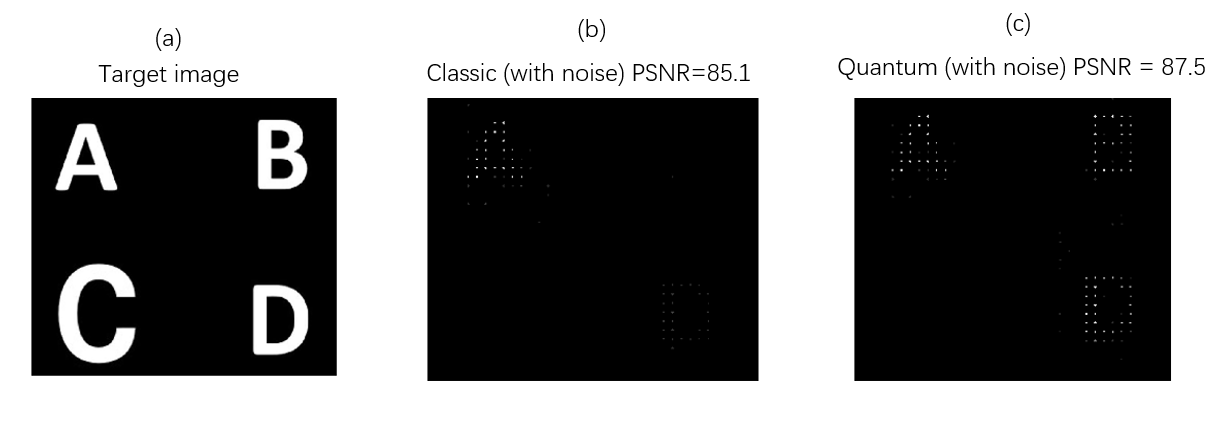}
	\caption{Classical and quantum peak signal-to-noise ratio (PSNR) in noisy conditions}
\end{figure}
Next, we present the classical PSNR and quantum PSNR under noise conditions for multiplexing holograms as shown in Fig.10. The classical PSNR is $85.1dB$. The quantum PSNR is $87.5dB$. Similarly, For multiplexing holograms, quantum holograms have a stronger noise resistance than that of the classical holograms.
Next, we respectively present the MSE, PSNR, and SSIM of classical BG holography, classical MBG holography, and quantum MBG holography under conditions with and without noise, as shown in Table 1.
\begin{table}[htbp]
	\centering
	\caption{Mean Squared Error (MSE), Peak Signal-to-Noise Ratio (PSNR), and Structural Similarity Index (SSIM) of Classical BG Holography, Classical MBG Holography, and Quantum MBG Holography under Noise-Free and Noisy Conditions}
	\scriptsize
\setlength{\tabcolsep}{0.8pt}
\renewcommand{\arraystretch}{1.5}
	\begin{tabular}{ccccc}
		\toprule
		Condition & Holography 
		 & MSE & PSNR & SSIM \\
       \toprule
		Noise-free & Classical BG & $1.2345\times10^{-4}$ & 87.2160 & 0.9903 \\
		Noise-free & Classical MBG & $2.9494\times10^{-4}$ & 93.4335 & 0.9887 \\
		Noise-free & Quantum MBG & $2.9467\times10^{-5}$ & 93.4375 & 0.9887 \\
		  \toprule
		Noisy & Classical BG   & $1.5088\times10^{-4}$ & 86.3444 & 0.9898 \\
		Noisy & Classical MBG & $1.0966\times10^{-4}$ & 87.7303 & 0.9911 \\
		Noisy  & Quantum MBG  & $2.9470\times10^{-5}$ & 93.4369 & 0.9887 \\
		\toprule
	\end{tabular}	
\end{table}

\section{Conclusion}
In this paper, we propose a high-dimensional quantum holography scheme based on multi-mode Bessel-Gaussian (MBG) orbital angular momentum (OAM) modes. We utilize spontaneous parametric down-conversion to generate OAM-entangled photon pairs and construct MBG-OAM quantum selective holograms by using the conical parameters and topological charges of the idler photons as joint encoding degrees of freedom. Image reconstruction is then achieved through correlated decoding of the corresponding mode parameters in the signal photons. Compared to the traditional holography schemes that rely solely on a single topological charge for encoding, our work introduce the conical parameter as an additional degree of freedom, thereby we extend the encoding space from a single-dimensional OAM mode to a higher-dimensional composite modes, significantly enhancing the system's multiplexing dimension, information capacity, and mode selectivity. The image reconstruction performance is subsequently validated for three scenarios: single-mode, dual-mode superposition, and multi-channel multiplexing. In order to quantitatively evaluating the quality of image reconstruction, we analysis the reconstruction results under different modes by using MSE, PSNR, and SSIM. Results show that the quantum MBG-OAM holography scheme can maintain a high image fidelity in both single-channel and multiplexed-channel cases. Notably, when noise is added, the quantum scheme demonstrates a higher PSNR compared to classical holography methods, indicating a more pronounced advantage in resisting noise interference. Our research provides a new approach for high-capacity, high-security quantum holography and establish a theoretical foundation for its applications in quantum imaging, high-dimensional quantum communication, and quantum information storage.

\textbf{Funding.} 
Financial support from the project funded by the State Key Laboratory of Quantum Optics and Quantum Optics Devices, Shanxi University, Shanxi, China (Grants No. KF202503). Zhejiang Key Laboratory of Quantum State Control and Optical Field Manipulation, Hangzhou Dianzi University, Hangzhou, China(Grants No.KYZ074326001).\\

\textbf{Disclosures.} The authors declare no conflicts of interest.\\

\textbf{Data availability statement.} Data underlying the results presented in this paper are not publicly available at this time but may be obtained from the authors upon reasonable request.\\

\bibliography{main}

@article{1,
	title={A new microscopic principle.},	
    author={Gabor, Dennis},
    journal={Nature},
    volume={161},
	pages={777},
	year={1948},
}

@article{2,
	title={Three-dimensional holographic image sensing and integral imaging display},
	author={Javidi, Bahram and Hong, Seung-Hyun},
	journal={Journal of Display Technology},
	volume={1},
	number={2},
	pages={341},
	year={2005},
	publisher={OSA}
}

@book{3,
	title={Holographic data storage},
	author={Coufal, Hans J and Psaltis, Demetri and Sincerbox, Glenn T and others},
	volume={8},
	year={2000},
	publisher={Springer}
}

@article{4,
	title={Securing information with optical technologies},
	author={Javidi, Bahram},
	journal={Physics Today},
	volume={50},
	number={3},
	pages={27-32},
	year={1997},
	publisher={AIP}
}

@inproceedings{5,
	title={Digital holography: methods and applications},
	author={Kreis, Thomas M and Jueptner, Werner PO and Geldmacher, Juergen},
	booktitle={International Conference on Applied Optical Metrology},
	volume={3407},
	pages={169-177},
	year={1998},
	organization={SPIE}
}

@incollection{6,
	title={Digital holography},
	author={Schnars, Ulf and Falldorf, Claas and Watson, John and J{\"u}ptner, Werner},
	booktitle={Digital Holography and Wavefront Sensing: Principles, Techniques and Applications},
	pages={39-68},
	year={2014},
	publisher={Springer}
}

@article{7,
	title={Three-dimensional imaging and processing using computational holographic imaging},
	author={Frauel, Yann and Naughton, Thomas J and Matoba, Osamu and Tajahuerce, Enrique and Javidi, Bahram},
	journal={Proceedings of the IEEE},
	volume={94},
	number={3},
	pages={636-653},
	year={2006},
	publisher={IEEE}
}

@article{8,
	title={Optical image encryption based on interference of polarized light},
	author={Zhu, Nan and Wang, Yongtian and Liu, Juan and Xie, Jinghui and Zhang, Hao},
	journal={Optics Express},
	volume={17},
	number={16},
	pages={13418-13424},
	year={2009},
	publisher={OSA}
}

@book{9,
	title={Holographic data storage: from theory to practical systems},
	author={Curtis, Kevin and Dhar, Lisa and Hill, Adrian and Wilson, William and Ayres, Mark},
	year={2011},
	publisher={John Wiley \& Sons}
}

@article{10,
	title={Dynamic characterization of MEMS diaphragm using time averaged in-line digital holography},
	author={Singh, Vijay Raj and Miao, Jianmin and Wang, Zhihong and Hegde, Gopalkrishna and Asundi, Anand},
	journal={Optics Communications},
	volume={280},
	number={2},
	pages={285-290},
	year={2007},
	publisher={Elsevier}
}

@article{11,
	title={Orbital angular momentum of light and the transformation of Laguerre-Gaussian laser modes},
	author={Allen, Les and Beijersbergen, Marco W and Spreeuw, RJC and Woerdman, JP},
	journal={Physical Review A},
	volume={45},
	number={11},
	pages={8185},
	year={1992},
	publisher={APS}
}

@article{12,
	title={Digital recording and numerical reconstruction of holograms},
	author={Schnars, Ulf and J{\"u}ptner, Werner PO},
	journal={Measurement Science and Technology},
	volume={13},
	number={9},
	pages={R85-R101},
	year={2002}
}

@article{13,
	title={Application of digital holography for nondestructive testing and metrology: a review},
	author={Kreis, Thomas},
	journal={IEEE Transactions on Industrial Informatics},
	volume={12},
	number={1},
	pages={240-247},
	year={2015},
	publisher={IEEE}
}

@article{14,
	title={Digital holographic three-dimensional video displays},
	author={Onural, Levent and Yara{\c{s}}, Fahri and Kang, Hoonjong},
	journal={Proceedings of the IEEE},
	volume={99},
	number={4},
	pages={576-589},
	year={2011},
	publisher={IEEE}
}

@article{15,
	title={Metasurface orbital angular momentum holography},
	author={Ren, Haoran and Briere, Gauthier and Fang, Xinyuan and Ni, Peinan and Sawant, Rajath and H{\'e}ron, S{\'e}bastien and Chenot, S{\'e}bastien and V{\'e}zian, St{\'e}phane and Damilano, Benjamin and Br{\"a}ndli, Virginie and others},
	journal={Nature Communications},
	volume={10},
	number={1},
	pages={2986},
	year={2019},
	publisher={Nature Publishing Group UK London}
}

@article{16,
	title={Terabit free-space data transmission employing orbital angular momentum multiplexing},
	author={Wang, Jian and Yang, Jeng-Yuan and Fazal, Irfan M and Ahmed, Nisar and Yan, Yan and Huang, Hao and Ren, Yongxiong and Yue, Yang and Dolinar, Samuel and Tur, Moshe and others},
	journal={Nature Photonics},
	volume={6},
	number={7},
	pages={488-496},
	year={2012},
	publisher={Nature Publishing Group UK London}
}

@article{17,
	title={Terabit-scale orbital angular momentum mode division multiplexing in fibers},
	author={Bozinovic, Nenad and Yue, Yang and Ren, Yongxiong and Tur, Moshe and Kristensen, Poul and Huang, Hao and Willner, Alan E and Ramachandran, Siddharth},
	journal={Science},
	volume={340},
	number={6140},
	pages={1545-1548},
	year={2013},
	publisher={AAAS}
}

@article{18,
	title={Orbital angular momentum holography for high-security encryption},
	author={Fang, Xinyuan and Ren, Haoran and Gu, Min},
	journal={Nature Photonics},
	volume={14},
	number={2},
	pages={102-108},
	year={2020},
	publisher={Nature Publishing Group UK London}
}

@article{19,
	title={High-capacity and multi-dimensional orbital angular momentum multiplexing holography},
	author={Zhang, Nian and Xiong, Baoxing and Zhang, Xiang and Yuan, Xiao},
	journal={Optics Express},
	volume={31},
	number={20},
	pages={31884-31897},
	year={2023},
	publisher={Optica Publishing Group}
}

@article{20,
	title={State of the art in holographic displays: a survey},
	author={Yara{\c{s}}, Fahri and Kang, Hoonjong and Onural, Levent},
	journal={Journal of Display Technology},
	volume={6},
	number={10},
	pages={443-454},
	year={2010},
	publisher={OSA}
}

@article{21,
	title={Bessel-Gaussian beam-based orbital angular momentum holography},
	author={Ji, Jiaying and Zheng, Zhigang and Zhu, Jialong and Wang, Le and Wang, Xinguang and Zhao, Shengmei},
	journal={Chinese Physics B},
	volume={33},
	number={1},
	pages={014204},
	year={2024},
	publisher={Chinese Physical Society and IOP Publishing Ltd}
}

@article{22,
	title={Simple technique for generating the perfect optical vortex},
	author={Garc{\'\i}a-Garc{\'\i}a, Joaqu{\'\i}n and Rickenstorff-Parrao, Carolina and Ramos-Garc{\'\i}a, Rub{\'e}n and Arriz{\'o}n, V{\'\i}ctor and Ostrovsky, Andrey S},
	journal={Optics Letters},
	volume={39},
	number={18},
	pages={5305-5308},
	year={2014},
	publisher={OSA}
}

@article{23,
	title={Holographic-inspired multiple circularly polarized vortex-beam generation with flexible topological charges and beam directions},
	author={Karimipour, Majid and Komjani, Nader and Aryanian, Iman},
	journal={Physical Review Applied},
	volume={11},
	number={5},
	pages={054027},
	year={2019},
	publisher={APS}
}

@article{24,
	title={High-dimensional entanglement-enabled holography},
	author={Kong, Ling-Jun and Sun, Yifan and Zhang, Furong and Zhang, Jingfeng and Zhang, Xiangdong},
	journal={Physical Review Letters},
	volume={130},
	number={5},
	pages={053602},
	year={2023},
	publisher={APS}
}

@article{25,
  title = {Measurement of the Spiral Spectrum of Entangled Two-Photon States},
  author = {Di Lorenzo Pires, H. and Florijn, H. C. B. and van Exter, M. P.},
  journal = {Physical Review Letters},
  volume = {104},
  number = {4},
  pages = {020505},
  year = {2010},
  publisher = {APS}
}

\end{document}